# Speed Tracking of a Linear Induction Motor - Enumerative Nonlinear Model Predictive Control

Jean Thomas and Anders Hansson, *Senior Member, IEEE*

*Abstract*— Direct torque control is considered as one of the most efficient techniques for speed and/or position tracking control of induction motor drives. However, this control scheme has several drawbacks: the switching frequency may exceed the maximum allowable switching frequency of the inverters, and the ripples in current and torque, especially at low speed tracking, may be too large. In this paper we propose a new approach that overcomes these problems. The suggested controller is a model predictive controller which directly controls the inverter switches. It is easy to implement in real time and it outperforms all previous approaches. Simulation results show that the new approach has as good tracking properties as any other scheme, and that it reduces the average inverter switching frequency about 95% as compared to classical direct torque control.

## I. INTRODUCTION

Nowadays, Linear Induction Motors (LIMs) are widely used in a variety of applications like transportation, conveyor systems, material handling, pumping of liquid metal, sliding door closers, robot base movers, office automation, drop towers, elevators etc., [1-2]. This is attributed to several advantages that the LIM posses, such as high starting thrust, alleviation of gears between motor and the motion devices, simple mechanical construction, no backlash and small friction, and suitability for both low speed and high speed applications, [3-5].

The driving principles of the LIM are similar to those of the traditional rotary induction motor. However, the control characteristics of the LIM are more complicated. This is attributed to the change in operating conditions due to mover speed, temperature, and rail configuration. Moreover, there are uncertainties existing in practical applications of the LIM which are usually composed of unpredictable plant parameter variations, external load disturbances, and unmodeled and nonlinear dynamics. Therefore, the design of LIM drive system should provide high tracking performance, and high dynamic stiffness to overcome the above challenges, [6-8].

J. Thomas is with the Division of Automatic Control, Beni-Suef University, Beni-Suef, Egypt (corresponding author phone: +20-125263476; fax: +46-13282622; e-mail: jthomas@mhe-spu.org).
A. Hansson is with the Division of Automatic Control, Department of Electrical Engineering, Linköping University, SE-58183 Linköping, Sweden (e-mail: hansson@isy.liu.se).

Several control techniques have been used to control the speed and/or position of induction motor drives. Among these control techniques, the method of Direct Torque Control (DTC) is considered as one of the most efficient techniques that can be used for induction motors, [9]. The basic characteristic of DTC is that the positions of the inverter switches are manipulated directly. The advantages of the DTC strategy are fast transient response, simple configuration, and high robustness against parameter variations. However, classical DTC has inherent drawbacks such as variable switching frequency, high torque and current ripples, high noise level at low speeds and also problems with the control of torque and flux at low speeds.

Model Predictive Control (MPC) has been applied to LIM drives for tracking of speed reference trajectories, [10]. Based on a linearized model of the LIM, the MPC controller calculates the optimal primary voltages while respecting constraints on flux and current in order to keep them within permissible values. It has been shown that the response is very fast as compared to classical DTC, and with almost no ripples in the current and torque signals. Moreover, it has been shown to be more robust against parameter uncertainty and load disturbance at high speed as well as at low speed. The MPC controller is used in conjunction with a PWM inverter. This often results in a high switching frequency at the inverter switches. Moreover, the computational burden of the on-line optimization and linearization makes real-time implementation impossible.

A MPC strategy for induction motor control based on feasibility and not on optimality is presented in [11]. The main objective there is to find a control input that keeps the controlled variables within their bounds, and select among the set of feasible control inputs the one that has minimum switching frequency. The motor together with the inverter are modeled as a hybrid system on so-called MLD-form, where the inverter switch positions are represented as integer variables. A performance improvement in terms of a reduction of the switching frequency as compared to classical DTC is shown. However, the approach provides only a feasible solution and no optimal solution. Moreover, the reformulation of the system into MLD-form and computing an explicit solution using a multi-parametric approach is computationally very demanding. Because of this only the case of a fixed operating point is considered. In [12] another MPC scheme is proposed that keeps the motor torque and the stator flux within given hysteresis bounds while minimizing the switching frequency of the inverter. The proposed Model Predictive DTC (MPDTC) scheme reduces the switching frequency by up to 50% as compared to other techniques, while respecting the torque and flux hysteresis bounds. In this approach the rotor speed dynamics are neglected and the speed is assumed to remain constant within the prediction horizon. A review of the most important types of predictive control used in power electronics and drives is presented in [13].

Just as in [12] we also propose to use MPC for control of an LIM. However, our approach employs an enumerative optimization of the MPC criterion function. With this approach we avoid any advanced modeling such as transforming the system to MLD-form. Moreover, we may consider the nonlinear dynamics and do not have to linearize the model. We will call our control strategy Enumerative Nonlinear MPC (ENMPC). Because the optimization is enumerative and over a small number of discrete variables, it is extremely fast, and hence admits real-time implementation. ENMPC is similar to the control scheme presented in [14]. There a predictive strategy for current control of a three-phase neural-point-clamped inverter is presented, where the behavior of the system is predicted for each possible switching position of the inverter, and the position that minimizes a given cost function is selected. Hence this approach is also enumerative. Several similar approaches can be found in [13]. However, they all consider a prediction horizon of one. In our work the prediction horizon is longer.

The paper is organized as follows: Section 2 briefly presents the dynamic model of the LIM. In section 3 the ENMPC controller is presented. The system configuration is described in Section 4. Simulation results and general remarks are presented in Section 5. Finally, conclusions and suggestions for future work are given in Section 6.

## II. Linear Induction Motor

### A. Dynamic Model of the LIM

The dynamic model of the LIM is similar to the traditional model of a three phase, Y-connected induction motor in $\alpha - \beta$ stationary frame, and it can be described by the following differential equations [15-17]:

$$\frac{d(i_{\alpha s})}{dt} = -\left(\frac{R_s}{\sigma L_s} + \frac{1-\sigma}{\sigma T_r}\right)i_{\alpha s} + \frac{Lm}{\sigma L_s L_r T_r}\lambda_{\alpha r} + \frac{n_p L_m \pi}{\sigma L_s L_r h}\upsilon\lambda_{\beta r} + \frac{1}{\sigma L_s}V_{\alpha s} \quad (1)$$

$$\frac{d(i_{\beta s})}{dt} = -\left(\frac{R_s}{\sigma L_s} + \frac{1-\sigma}{\sigma T_r}\right)i_{\beta s} - \frac{n_p L_m \pi}{\sigma L_s L_r h}\upsilon\lambda_{\alpha r} + \frac{Lm}{\sigma L_s L_r T_r}\lambda_{\beta r} + \frac{1}{\sigma L_s}V_{\beta s} \quad (2)$$

$$\frac{d(\lambda_{\alpha r})}{dt} = \frac{L_m}{T_r}i_{\alpha s} - \frac{1}{T_r}\lambda_{\alpha r} - \frac{n_p \pi}{h}\upsilon\lambda_{\beta r} \quad (3)$$

$$\frac{d(\lambda_{\beta r})}{dt} = \frac{L_m}{T_r}i_{\beta s} + \frac{n_p \pi}{h}\upsilon\lambda_{\alpha r} - \frac{1}{T_r}\lambda_{\beta r} \quad (4)$$

$$\frac{d(\upsilon)}{dt} = \frac{1}{M}F_e - \frac{D}{M}\upsilon - \frac{1}{M}F_L \quad (5)$$

where $T_r = \frac{L_r}{R_r}$, $\sigma = 1 - \frac{L^2_m}{L_s L_r}$, and where

- $D$ : viscous friction and iron-loss coefficient,
- $F_L$ : external force disturbance,
- $L_s$ : primary inductance per phase
- $F_e$ : electromagnetic force,
- $h$ : pole pitch,
- $L_r$ : secondary inductance per phase,

$L_m$ : magnetizing inductance per phase,

$n_p$ : number of pole pairs.

$R_r$ : secondary resistance per phase,

$\upsilon$ : mover linear velocity,

$i_{\alpha s}, i_{\beta s}$ : $\alpha - \beta$ primary current components,

$\sigma$ : leakage coefficient.

$M$ : total mass of the moving element,

$R_s$ : primary winding resistance per phase,

$T_r$ : secondary time constant,

$\lambda_{\alpha r}, \lambda_{\beta r}$ : $\alpha - \beta$ secondary flux components,

$V_{\alpha s}, V_{\beta s}$ : $\alpha - \beta$ primary voltage components,

The electromagnetic force can be described in the $\alpha - \beta$ fixed frame as:

$$F_e = k_f \left( \lambda_{\alpha r} i_{\beta s} - \lambda_{\beta r} i_{\alpha s} \right) \tag{6}$$

where $k_f$ is the force constant which is equal to:

$$k_f = \frac{3 n_p L_m \pi}{2 L_r h} \tag{7}$$

We will use a forward-Euler discretization of the nonlinear differential equations to obtain a discrete-time model suitable for our purposes of MPC.

## B. DC-AC Inverter

The three phase two-level DC-AC inverter used to drive the LIM is shown in Figure 1. The three switches can be modeled by three binary variables $u_{1,2,3} \in \{0,1\}$ representing on/off positions, which imply the following relation:

$$u_i = \begin{cases} 1 & \Leftrightarrow \quad v_i = \frac{V_{dc}}{2} \\ 0 & \Leftrightarrow \quad v_i = -\frac{V_{dc}}{2} \end{cases} \quad i = 1,2,3 \tag{8}$$

where $V_{dc}$ is the DC voltage source. The three switches have 8 possible different position combinations. The relation between the primary voltage components $V_{\alpha s}, V_{\beta s}$ and the switching positions are given by the following equation:

$$\begin{bmatrix} V_{\alpha s} \\ V_{\beta s} \end{bmatrix} = V_{dc} \times \frac{3}{2} \begin{bmatrix} \frac{2}{3} & -\frac{1}{3} & -\frac{1}{3} \\ 0 & -\frac{1}{\sqrt{3}} & \frac{1}{\sqrt{3}} \end{bmatrix} \begin{bmatrix} u_1 \\ u_2 \\ u_3 \end{bmatrix} \tag{9}$$

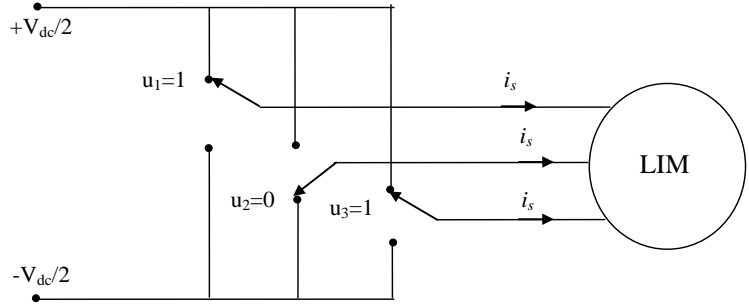

Figure 1: Three-phase inverter driving the LIM.

*C. Control Objectives*

The main objective is to control the speed of the LIM drive to track the given speed reference. The controller controls the three inverter switching positions to provide the necessary primary voltage to track the speed reference. It is recommended to minimize the average switching frequency of the inverter switches. Constraints over secondary flux and primary current should be considered: the secondary flux should be less than 0.45 Wb, and the primary current should be less than 50A.

III. ENUMERATIVE NONLINEAR MPC CONTROLLER

The main idea of MPC is to use a model of the plant to predict future outputs of the system. Based on this prediction, at each sampling period, a sequence of future control values is computed through an on-line optimization process, which maximizes the tracking performance while satisfying constraints. Only the first value of this optimal sequence is applied to the plant, and the whole procedure is repeated again at the next sampling period according to what usually is called a 'receding' horizon strategy, [18-20].

Applying MPC to an LIM grants a better performance than the classical DTC approach, [10], but the main drawbacks of this technique are the heavy on-line computations that make it inapplicable in real-time, and also the high switching frequency that may exceed the maximum allowable frequency. Because of these drawbacks, the following ENMPC controller is proposed.

As the three switches of the inverter have only eight different position combinations, an analytical computation of the tracking performance, for the eight possible position combinations can be performed. Then the position of the switches, which are the manipulated variables, that maximizes the tracking performance is selected. The eight different combinations of positions are elements of the set

$$U = \{(1,0,0), (0,1,0), (0,0,1), (1,1,0), (0,1,1), (1,0,1), (0,0,0), (1,1,1)\} \tag{10}$$

The objective function that captures the tracking performance includes the error between the actual speed and the speed reference trajectory. To minimize the inverter switching frequency a penalty term on the control variations is

included in the objective function. The considered objective function is:

$$J = \sum_{j=1}^{N} Q(\hat{v}(k+j|k) - w(k+j))^2 + \sum_{j=0}^{N_u-1} P_j T(u(k+j), u(k+j-1)) \qquad (11)$$

where $\hat{v}$ is the predicted future speed, $w$ is the speed reference, $u$ is the control signal, and where $Q$ and $P_j$ are positive constants. The second term which penalizes the input switching, measures directly the switching number, i.e. $T(u(k+j), u(k+j-1))$ is the number of switches as defined in Table 1. The value in row $i$ and column $j$ is showing the number of switches when $u(k+j-1)$ has the value of element $i$ and $u(k+j)$ has the value of element $j$ in $U$ in (10). The objective function (11) is minimized subject to constraints that describe the discretized dynamics in (1)-(9).

Table 1: Number of control switches, where indexes 1,2, …, 8 refers to elements in $U$ in (10)

| $u(k+j-1)$ \ $u(k+j)$ | 1 | 2 | 3 | 4 | 5 | 6 | 7 | 8 |
|---|---|---|---|---|---|---|---|---|
| 1 | 0 | 2 | 2 | 1 | 3 | 1 | 1 | 2 |
| 2 | 2 | 0 | 2 | 1 | 1 | 3 | 1 | 2 |
| 3 | 2 | 2 | 0 | 3 | 1 | 1 | 1 | 2 |
| 4 | 1 | 1 | 3 | 0 | 2 | 2 | 2 | 1 |
| 5 | 3 | 1 | 1 | 2 | 0 | 2 | 2 | 1 |
| 6 | 1 | 3 | 1 | 2 | 2 | 0 | 2 | 1 |
| 7 | 1 | 1 | 1 | 2 | 2 | 2 | 0 | 3 |
| 8 | 2 | 2 | 2 | 1 | 1 | 1 | 3 | 0 |

The constants $P_j$ should impose more penalties over the first time-steps than the later steps, to force the transition of the switches to occur as late as possible [21-22]. This is accomplished by the following constraints:

$$P_0 > P_1 > \cdots > P_{N_u-1}. \qquad (12)$$

Elimination of small steady-state errors can be accomplished in different ways. The method we have used involves modifying the objective function to not only minimize the tracking error but to also minimize a sum of old tracking errors. Thus the objective function (11) is redefined as:

$$J = \sum_{j=1}^{N} Q(\hat{v}(k+j|k) - w(k+j))^2 + \sum_{j=1}^{N} P_i(\hat{E}(k+j))^2 + \sum_{j=0}^{N_u-1} P_j T(u(k+j), u(k+j-1)) \qquad (13)$$

Here $\hat{E}(k)$ is a prediction of the sum of the tracking error $E(k)$, where $E(k)$ is defined as follows:

$$E(k+1) = E(k) + K(w(k) - v(k)) \qquad (14)$$

where $v$ is the measured speed, $w$ is the speed reference, and where $K$ is a gain. To avoid that $E$ becomes too large we may replace (14) with $E(k+1) = E(k)$ when $|E(k)|$ is larger than a certain limit.

Our method is aiming at providing integral control. There are other approaches to eliminate steady-state offset, see e.g. [23]. Notice that many of the traditional integration methods are not possible to use in this application where the control signal is not a continuous valued signal.

The concept of control horizon ($N_u < N$) is used to reduce the number of decision variables and thus the computational time. Other methods to reduce the number of optimization variables could also have been used, e.g. blocking of the input variables technique, [24]. The objective function (13) is evaluated $s = 8^{N_u}$ times at each time step, and the first control signal in the sequence $u^{opt} = (u(k),...,u(k+N_u-1))$ corresponding to the minimum objective function value is then selected and applied to the inverter switches.

Increasing the prediction horizon $N$ will lead to more accurate choice of control signals. However, increasing the prediction horizon will increase the computational time. To account for that, we propose to use different discrete time models with different sampling times as described in [19]. For the first sampling steps we use a motor model with the true sampling time, and then for later sampling steps we use another model with longer sampling time. This will increase the prediction interval with less number of prediction steps as compared to when using the same sampling time for all predictions.

To avoid examining all possible input combinations over the control horizon $N$ the following incremental algorithm is proposed to compute the optimal control signal sequence. Here $u^i$ is a candidate optimal control signal sequence that is an element in $U \times U \times ... \times U$, where the number of Cartesian products is $s$-1.

*Algorithm 1*

1- Initializing with $J_{opt} = \infty$, $J^i(k) = 0$

2- for $i = 1 : s$

   3- for $j = 1 : N$

      4- let

$$J^i(k+j) = J^i(k+j-1) + f\left(\hat{v}(k+j|k), u^i(k+j-1)\right)$$ where $f\left(\hat{v}(k+j|k), u^i(k+j-1)\right)$ is the incremental cost at time $k+j$ due to the control signal $u^i(k+j-1)$.

      5- If $J^i(k+j) > J_{opt}$

         break and go to step 2

      end if

   end for

   6- At $j = N$

if $J^i(k+N) < J_{opt}$, $J_{opt} := J^i(k+N)$

end if

end for

7- $J_{opt}^* = J_{opt}$ the optimal value

The incremental cost (in step 4 of Algorithm 1) is the predicted cost at time step $k+j$ due to the control signal $u^i(k+j-1)$, and it is given by

$$f(\hat{v}(k+j|k), u^i(k+j-1)) = Q(\hat{v}(k+j|k) - w(k+j))^2 + P_i(\hat{E}(k+j|k))^2 + P_j T(u(k+j), u(k+j-1)).$$

Algorithm 1 stops the cost function calculations for the control sequence $u^i$ prematurely if the cost function at prediction step $j$ is higher than the current upper bound $J_{opt}$. This saves computational time. The algorithm is similar to one of the pruning rules in the Branch and Bound (BB) algorithm for solving integer programs.

One of the main advantages of MPC is its ability to deal with constraints, i.e. offering optimal control while respecting the given constraints. Including flux and current constraints into our proposed controller is simple. As an example a maximum flux constraint can be obtained simply by adding the following line to the controller code:

- If $\lambda_r(i_s) > \lambda_{r\max}(i_{s\max})$, $J^i(k+j) := \infty$.

In a similar way any switching position combination that leads to violation of a current and/or the flux constraints may be avoided. These constraints will allow the controller to track the speed reference, which is the main objective, while adjusting the flux and the current within their constraints.

The proposed controller is faster than other standard techniques for solving integer programming problems like for example BB, which is generally considered as one of the most effective techniques. At each step in the BB algorithm a relaxed optimization problem, often a convex quadratic program, is solved where a certain number of integer variables is relaxed to continuous variables with values constrained in [0,1]. Solving these relaxed optimization problems takes more time than the analytical computation of the objective function when the number of optimization variables is small, which is the case in the application in this paper. Moreover, the relaxed problem for the MPC controller we suggest would not be a quadratic program, since we have introduced a penalty term on the number of switches and because the discretized dynamics of the LMI is not linear. Hence they can be expensive to solve.

The advantages of the proposed technique besides its simple design and implementation are that there is no complicated on-line optimization to be performed. Furthermore, there is no need to linearize the LIM model as was necessary in [10]. Moreover there is no need to reformulate the system in the hybrid system framework, neither as a piecewise affine model nor as an MLD model as done in [11]. Operating point changes are also easily incorporated in our framework. Even preview control is possible, i.e. in case the future value of the reference value is known, this can be taken into account. There are several applications of LIMs where this is potentially advantageous, e.g. elevators and autonomous trains.

The developed technique significantly reduces the computational time. Moreover, one extra dimension of freedom through the choice of the weights $P_j$ has been added, which enables a trade-off between the average switching frequency and the speed tracking performance. Note that reducing the torque ripple can only be achieved by increasing the switching frequency and vice versa, [25].

## IV. SYSTEM CONFIGURATION

A block diagram of the linear induction motor controlled with the proposed ENMPC controller is shown in Figure 2. The system consists of the LIM, an inverter, the ENMPC controller, and a flux estimator. The input signals to the ENMPC controller are the speed reference $w$, the LIM velocity $v$, the primary currents $i_{\alpha s}$ and $i_{\beta s}$, and estimates of the secondary fluxes $\lambda_{\alpha r}$ and $\lambda_{\beta r}$.

The secondary flux components are estimated using the voltage and current signals as follows, [26-27]:

$$\lambda_{\alpha r} = (L_r / L_m)(\lambda_{\alpha s} - \sigma L_s i_{\alpha s}) \quad , \quad \lambda_{\beta r} = (L_r / L_m)(\lambda_{\beta s} - \sigma L_s i_{\beta s}) \tag{15}$$

where $\lambda_{\alpha s} = \int (V_{\alpha s} - i_{\alpha s} R_s) dt, \quad \lambda_{\beta s} = \int (V_{\beta s} - i_{\beta s} R_s) dt$

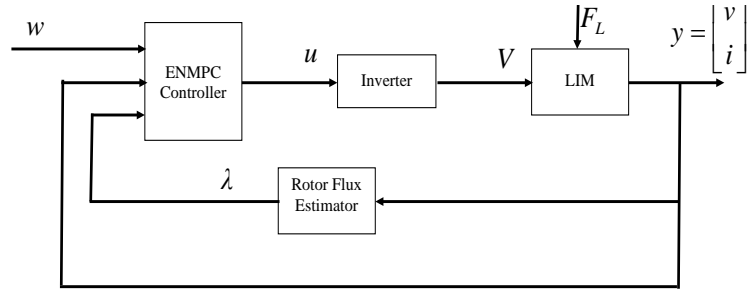

Figure 2. Block diagram of the LIM drive controlled with the proposed ENMPC controller.

### A. Control Configuration

Different values for the control horizon and the prediction horizon, $N_u$, $N$, respectively, have been considered. In experiments we have seen that for $T_s = 100\ \mu s$ the choice of $N_u = 2$, $N = 10$ provides a good performance at high and low speed tracking, and that there is no need to increase the control horizon. With a shorter control horizon $N_u = 1$, we only have a slightly lower performance at low speed.

After successive tuning iterations, the parameters of the MPC controller that give a good response are: control horizon $N_u = 1$, prediction interval $= 10 \times T_s$. The concept of multiple discrete models, as mentioned previously, is used to reduce the number of prediction steps; a model with sampling time $T_s$ is used for the first two steps, and then a model with sampling time equal $4T_s$ is used for the next 2 steps, i.e. the prediction interval of in total $10T_s$ is covered with 4 prediction steps. The weights in the objective function has been chosen as $P_j = 1$, $Q = 1000000$, $P_i = 500$ and $k = 150$.

The considered constraints on fluxes and currents force the controller to keep them within their minimum and maximum limits.

## V. RESULTS AND DISCUSSIONS

Computer simulations have been carried out in order to validate the proposed scheme. The Matlab/Simulink software package with a c-mex file interfacing the proposed controller has been used. Different operating conditions including load change and various speed trajectories have been considered. The data of the LIM used for simulations are, [28]: 3-phase, Y-connected, 8-pole, 3-kW, 60-Hz, 180-V, 14.2 A. The motor parameters are listed below in Table 2.

The nominal force for the considered motor is around 650N; it can be calculated roughly as follows: Nominal force = Force constant x rated current x rated flux.

TABLE 2. PARAMETERS AND DATA OF THE LIM

| Rs (Ω) | 5.3685 | Pole pitch, h (m) | 0.027 |
|---|---|---|---|
| Rr (Ω) | 3.5315 | Total mass of the mover, M (kg). | 2.78 |
| Ls (H) | 0.02846 | Viscous friction and iron-loss coefficient, D (kg/s) | 36.0455 |
| Lr (H) | 0.02846 | Force constant, Kf (N/wb.A) | 592 |
| Lm (H) | 0.02419 | Rated secondary flux, (wb) | 0.056 |

Figure 3 shows the speed responses of the proposed ENMPC controller and also that obtained with classical DTC; the DTC includes an outer speed control loop, which is a PI controller, to track the speed reference, [29]. The LIM is assumed to start at t=0 and accelerated up to 2 m/s in 0.2 seconds. Then the motor speed reference is kept constant at this value during the remaining simulation period. A load force is stepped from 350 N to 500 N at t = 0.5 seconds. It is worth to note that the acceleration period (0.2 s) is considered to be long enough for the motor to attain the desired speed (2 m/s). This is because the tested motor is of small size and weight and therefore has a mechanical time constant of 0.077 s.

While the DTC needs about 0.1 seconds in order to attain the steady state value both from start and after the load disturbance takes place, the ENMPC controller needs only around 0.01 seconds , and it also has much better tracking performance. At t=0.5 seconds there is a very small dip in the speed response due to the load change, but the controller succeeds in restoring the speed reference very quickly (a zoom of the load change response is shown in the lower part of Figure 3). It is obvious that the proposed controller response is much faster than that of the DTC response and that it is able to deal with load changes more efficiently.

For this simulation, the maximum number of transitions of the inverter switches for the DTC approach was 51014/s, while for the enumerative approach it was only 1542/s, which is a reduction of about 97% of the switching frequency.

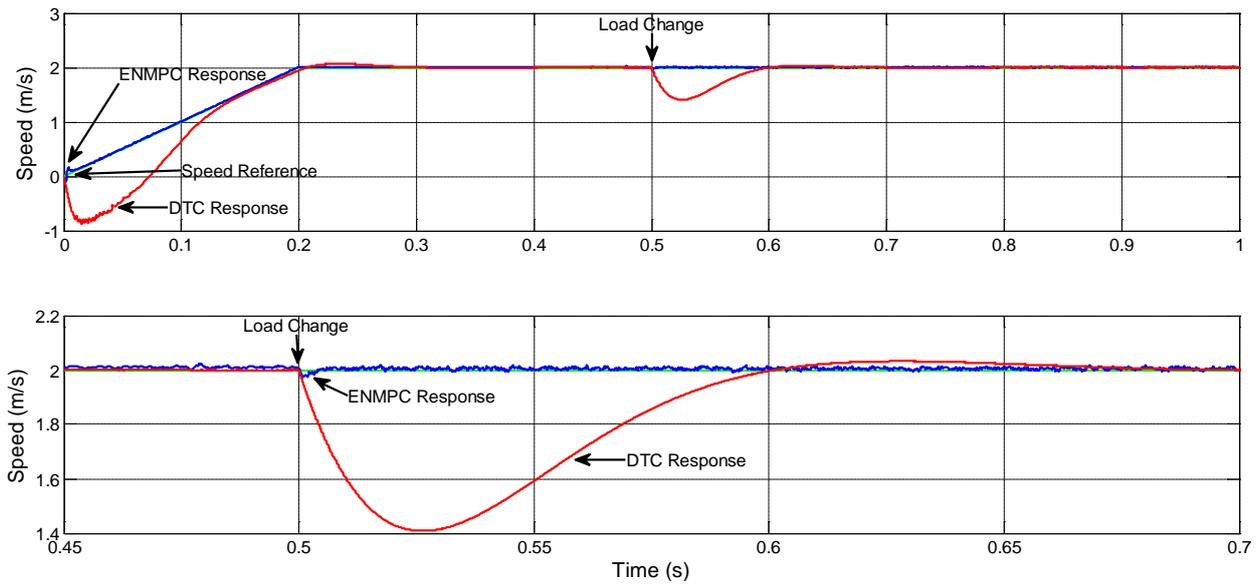

Figure 3. ENMPC response versus DTC response

The performance of the ENMPC controller was tested also at low speed (0.1 m/s), which is more challenging, with the same load change as in the previous scenario. The same controller parameters were used. The worst case of ripples for the current and torque occurs at such low speed. Figure 4 shows the simulation results. They are from top to bottom: the speed response, the developed electromagnetic force (Fe), the speed response zoomed around the load change together with the three phase primary currents. The lower two plots are zoom views to see in more detail what happens when the load change occurs. Again the controller responds quickly to the load disturbance and behaves well at low speed. The 3-phase currents and the electromagnetic force have much lower ripples as compared with the DTC technique. The maximum switching frequency in this case was 1890/s, which is a reduction of 98% as compared to DTC for the same scenario.

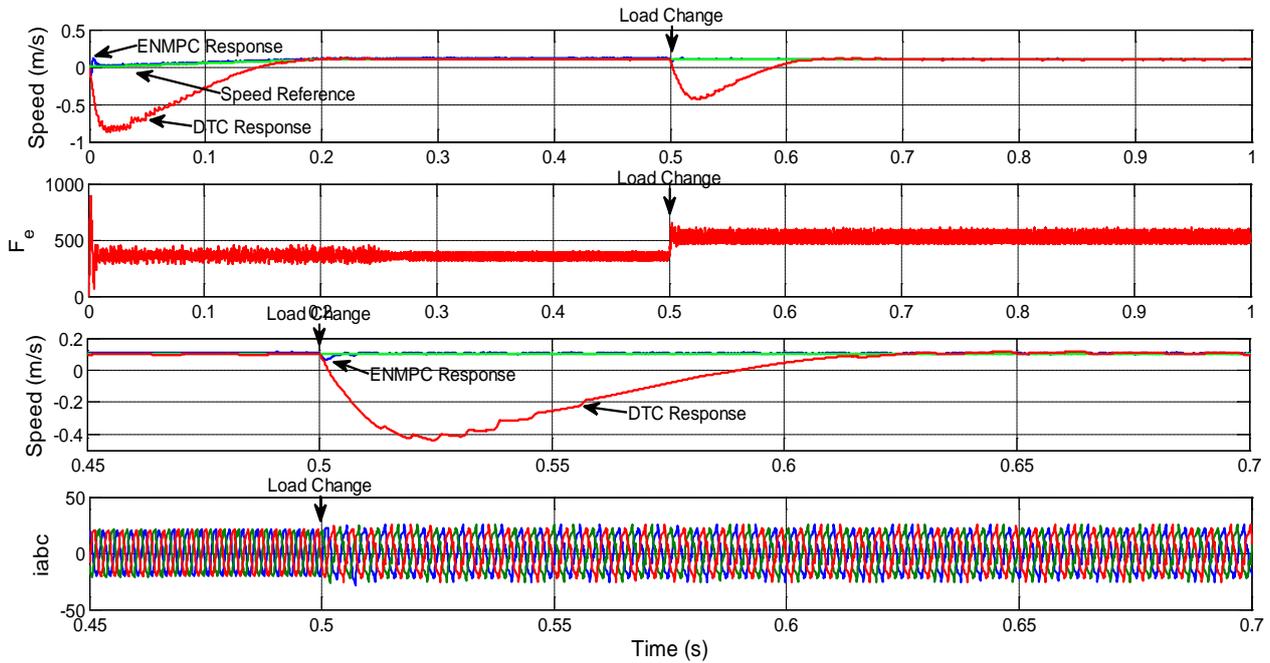

Figure 4. Simulation results obtained with the ENMPC controller at low speed.

The robustness of the ENMPC controller against parameter variations was also examined. The value of the primary resistance $R_s$ was changed in the LIM model but kept at its nominal value in the model based ENMPC controller. Figure 5 shows the ENMPC controller response for the case of $R_s$ increased by 50% (the upper part) and reduced by 50% (the lower part) at low speed, with load changes from 350N to 500N at $t = 0.5$ seconds. Figure 5 shows also the DTC response as well as the response of the MPC controller of [10] for the same scenarios. The reason for investigating this parameter change is that it has significant effect on the flux estimation at low speeds. As is the case for the MPC controller of [10], and the DTC, the ENMPC controller is robust against these parameter variations.

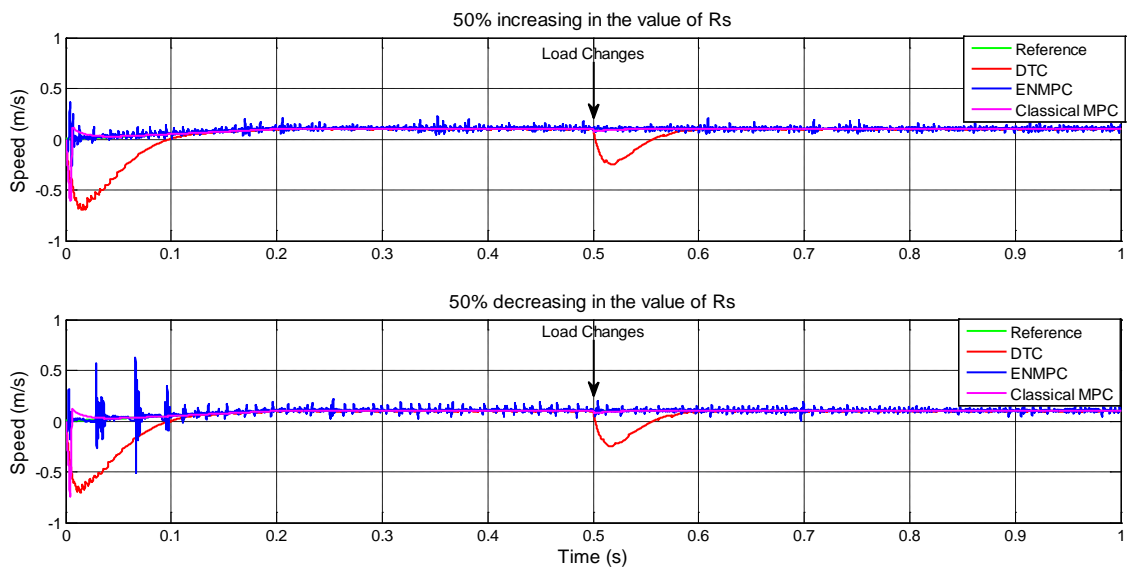

Figure 5. ENMPC, MPC of [10] and DTC responses for parameter variation; $R_s$ increased by 50% (upper part), and reduced by 50% (lower part), at low speed.

The choice of the weighting for the manipulated variable $P_j$ affects the switching frequency. For higher values of $P_j$, the penalties on the control variations increase which leads to a reduction of the switching frequency. However this could have a bad effect on the performance of speed tracking. A trade-off between the switching frequency and the speed tracking performance, can be achieved through the choice of $P_j$. Figure 6 shows the effect of the penalty matrix $P_j$ on the tracking performance for $P_j = 10000$ and $P_j = 1$. The same high-speed scenario is considered as before. With $P_j = 10000$ we have a lower switching frequency (919/s) than for $P_j = 1$, but a relatively poor performance with respect to ripple.

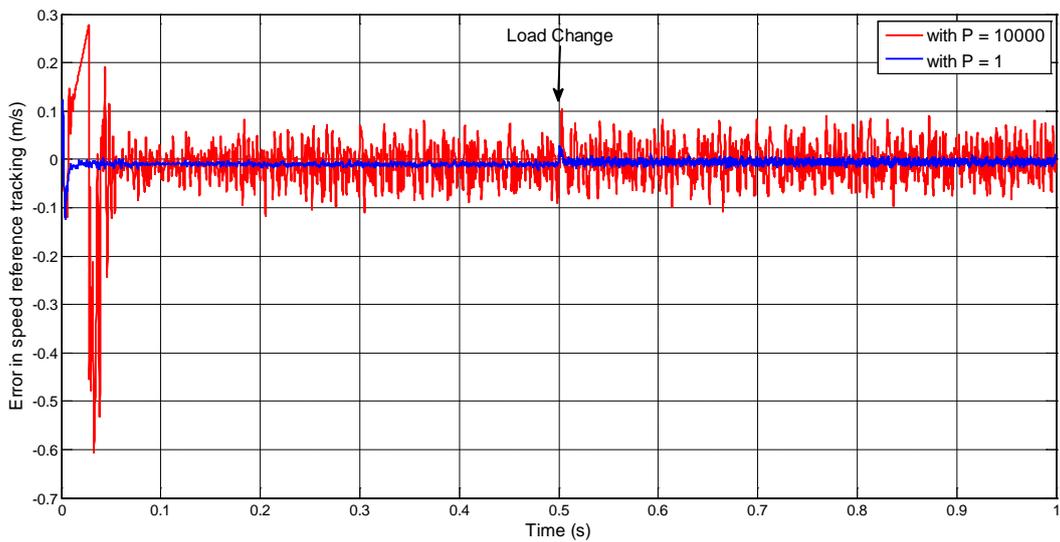

Figure 6. Effect of the penalty matrix $P_j$ on the tracking performance.

All constraints on fluxes and currents are satisfied in the previously investigated scenarios as shown for example in Figure 7, where typical constraints are: secondary flux less than 0.45 and primary current value less than 50 A.

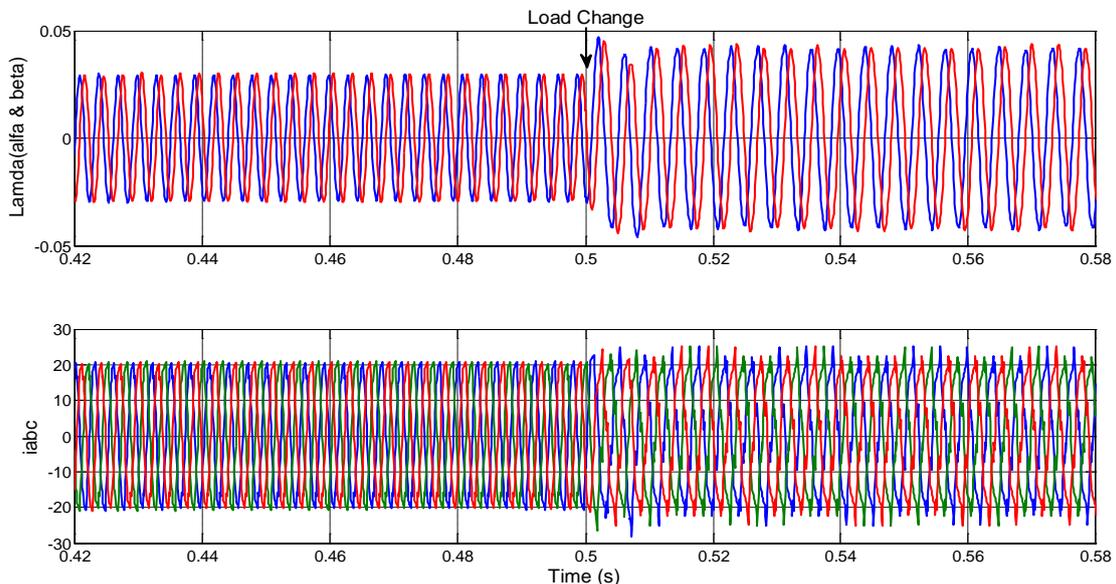

Figure 7. The secondary flux and primary currents are within the constraints

*A. Complexity analysis*

The computational time is mainly affected by the control horizon $N_u$, where the objective function is evaluated $8^{N_u}$ times at each sampling step. With the proposed technique with $N_u = 1$ and $N = 10$ and one discrete model it takes on average $120\ \mu s$ to compute the control signal. Applying the multiple model concept, i.e. 10 prediction intervals covered by 4 steps, reduces the computational time to $60\ \mu s$, which is less than the sampling time. The simulation results presented were done with the latter values. With a control horizon of $N_u = 2$ and multiple models as before the computational time would be $290\ \mu s$ instead. All the simulations have been run in Matlab 7.8 on a 3 MHz PC with 2 Mram.

Needless to say, for the classical MPC controller the computational time because of on-line linearization and optimization will be substantially higher, even if commercial optimization software such as CPLEX, [30], is used. This makes it impractical to implement in real time.

Previous simulation results prove the success of the presented technique; it has good performance with respect to speed tracking at high speed as well as at low speed. It is robust against load changes and parameter variations. It is successful in reducing the average switching frequency. Moreover, the proposed controller reduces significantly the computational time as compared to classical MPC approaches, which makes it applicable in real time application.

## VI. CONCLUSIONS

This paper considers speed tracking for a linear induction motor. It presents a new ENMPC controller based on the model predictive control approach. The developed controller controls directly the inverter switches to track the speed trajectory of the linear induction motor drive. The controller succeeds in tracking the speed trajectory at both high and low speed, and it reduces the switching frequency with about 95% as compared to classical DCT.

The proposed MPC controller response has many advantages; besides being simple to construct and to implement, it has a very fast response, lower ripples over currents and electromagnetic force in comparison to the DTC approach, and robustness against load changes and parameter variations. With this technique there is no need to use a PWM inverter, and moreover, it reduces significantly the computational time, which is an inherent drawback of classical MPC controllers. Thus real time implementation is possible.

Future work will include experimental works to validate this technique in practice. Finally, the same technique will be examined for other machines like rotary induction motors, and permanent magnet synchronous motors.